\documentclass[3p, twocolumn]{elsarticle}
\pdfoutput=1 

\usepackage[utf8x]{inputenc}
\usepackage{amssymb,amsmath, amsfonts}
       \usepackage{paralist}
       \usepackage{subfigure}
\usepackage[toc]{glossaries}
       \usepackage{paralist}

\def\fig{Fig.~}
\def\eq{Eq.~(} 
\begin{document}
%  \linenumbers

\begin{frontmatter}
\makeglossaries

\newacronym{UVB}{UVB}{Ultraviolet B}

\newacronym{CLOUDS}{CLOUDS}{Cosmics Leaving Outdoor Droplets}

\newacronym{eV}{eV}{electron Volt}

\newacronym{CERN}{CERN}{Organisation Europ\'{e}enne pour la Recherche Nucl\'{e}aire (European Organization for Nuclear Research)}

\newacronym{DNA}{DNA}{Deoxyribonucleic acid}

\newacronym{EF}{EF}{Exposure Facility}

\newacronym{CNRS}{CNRS}{Centre national de la recherche scientifique (French National Center for Scientific Research)}

 \newacronym{MIT}{MIT}{Massachusetts Institute of Technology}

\newacronym{JEM-EUSO}{JEM-EUSO}{Extreme Universe Space Observatory on-board the Japanese Experiment Module}

\newacronym{ASIC}{ASIC}{Application Specific Integrated Circuit}

\newacronym{SPACIROC}{SPACIROC}{Spatial Photomultiplier Array Counting and
Integrating Readout Chip}

\newacronym{TA}{TA}{Telescope Array}

\newacronym{Auger}{Auger}{Pierre Auger Observatory}

\newacronym{PMT}{PMT}{Photomultiplier Tube}

\newacronym{MAPMT}{MAPMT}{Multi-Anode Photomultiplier Tube}

\newacronym{DAQ}{DAQ}{Data Acquisition System}

\newacronym{RMS}{RMS}{Root Mean Square}

\newacronym{DAC}{DAC}{Digital-to-Analog Convertor}

\newacronym{SFR}{SFR}{Stellar Formation Rate}

\newacronym{FR-II}{FR-II}{Fanaroff-Riley Class II}

\newacronym{Mpc}{Mpc}{MegaParsec}

\newacronym{ODB}{ODB}{Online DataBase}

\newacronym{FSR}{FSR}{Full Scale Range}

\newacronym{LSB}{LSB}{Least Significant Bit}

\newacronym{DNL}{DNL}{Differential Non-Linearity}

\newacronym{INL}{INL}{Integral Non-Linearity}

\newacronym{NIM}{NIM}{Nuclear Instrumentation Module}

\newacronym{QDC}{QDC}{Charge-to-Digital Convertor}

\newacronym{ADC}{ADC}{Analog-to-Digital Convertor}

\newacronym{TDC}{TDC}{Time-to-Digital Convertor}

\newacronym{CAMAC}{CAMAC}{Computer Automated Measurement And Control}

\newacronym{PDM}{PDM}{Photodetection Module}

\newacronym{EC}{EC}{Elementary Cell}

\newacronym{LIDAR}{LIDAR}{Light Detection And Ranging}

\newacronym{CLF}{CLF}{Central Laser Facility}

\newacronym{ELS}{ELS}{Electron Light Source}

\newacronym{TALE}{TALE}{Telescope Array Low Energy Extension}

\newacronym{XLF}{XLF}{eXtreme Laser Facility}

\newacronym{AMIGA}{AMIGA}{Auger Muons and Infill for the Ground Array}

\newacronym{HEAT}{HEAT}{High Elevation Auger Telescopes}

\newacronym{AERA}{AERA}{Auger Engineering Radio Array}

\newacronym{IR}{IR}{Infrared}

\newacronym{PCB}{PCB}{Printed Circuit Board}

\newacronym{UV}{UV}{Ultraviolet}

\newacronym{CW-HVPS}{CW-HVPS}{Cockcroft-Walton High Voltage Power Supply}

\newacronym{HVPS}{HVPS}{High Voltage Power Supply}

\newacronym{PS}{PS}{Power Supply}

% \newacronym{pe}{PE}{PhotoElectron}
\newacronym{pe}{photoelectron}{photoelectron}

\newacronym{SNO}{SNO}{Sudbury Neutrino Observatory}

\newacronym{spe}{spe}{single photoelectron}

\newacronym{FD}{FD}{Fluorescence Detector}

\newacronym{FoV}{FoV}{Field of View}

\newacronym{SD}{SD}{Surface Detector}

\newacronym{SiPM}{SiPM}{Silicon Photomultiplier}

\newacronym{UHECR}{UHECR}{Ultra-High Energy Cosmic Ray}

\newacronym{EAS}{EAS}{Extensive Air Showers}

\newacronym{PMMA}{PMMA}{Poly(methyl methacrylate)}

\newacronym{BBM}{BBM}{Bread Board Model}

\newacronym{TLE}{TLE}{Transient Luminous Events}

\newacronym{HTV}{HTV}{H-IIB Transfer Vehicle}

\newacronym{TDRS}{TDRS}{Tracking and Data Relay Satellite}

\newacronym{JAXA}{JAXA}{Japan Aerospace Exploration Agency}

\newacronym{GLS}{GLS}{Global Light System}

\newacronym{PTT}{PTT}{Persistent Track Trigger}

\newacronym{LTT}{LTT}{Linear Track Trigger}

\newacronym{TCU}{TCU}{Telemetry Command Unit} 

\newacronym{IDAQ}{IDAQ}{Data Acquisition Interface}

\newacronym{MEM}{MEM}{MicroElectroMechanical}

\newacronym{DMSPac}{DMSP}{Defense Meteorological Satellite Program}

\newacronym{MDP}{MDP}{Mission Data Processor}

\newacronym{CCB}{CCB}{Cluster Control Board}

\newacronym{GLS-X}{GLS-X}{Global Light System - Xenon Flashers}

\newacronym{GLS-XL}{GLS-XL}{Global Light System - Xenon Flashers and Lasers}

\newacronym{FSU}{FSU}{Fast Shaper Unit}

\newacronym{NIST}{NIST}{National Institute of Standards and Technology}

\newacronym{LED}{LED}{Light-Emitting Diode}

\newacronym{VME}{VME}{Versa Module European bus}

\newacronym{ISS}{ISS}{International Space Station}

\newacronym{AMS}{AMS}{Atmospheric Monitoring System}

\newacronym{GTU}{GTU}{Gate Time Unit}

\newacronym{MIDAS-UK}{MIDAS-UK}{Multi Instance Data Acquisition System}

\newacronym{MIDAS}{MIDAS}{Maximum Integrated Data Acquisition System}

\newacronym{LHC}{LHC}{Large Hadron Collider}

\newacronym{HiRes}{HiRes}{High Resolution Fly's Eye}

\newacronym{CMB}{CMB}{Cosmic Microwave Background}

\newacronym{GRB}{GRB}{Gamma Ray Burst}

\newacronym{GPIB}{GPIB}{General Purpose Interface Bus}

\newacronym{USB}{USB}{Universal Serial Bus}

\newacronym{CCD}{CCD}{Charge-Coupled Device}

\newacronym{GZK}{GZK}{Greisen, Zatsepin, and Kuzmin}

\newacronym{EM}{EM}{Electromagnetic}

\newacronym{LPM}{LPM}{Landau-Pomeranchuk-Migdal}

\newacronym{DICE}{DICE}{Dual Imaging Cherenkov Experiment}

\newacronym{KASCADE}{KASCADE}{Karlsruhe Shower Core and Array Detector}

\newacronym{AGASA}{AGASA}{Akeno Giant Air Shower Array} 

\newacronym{UHE}{UHE}{Ultra High Energy}

\newacronym{CNES}{CNES}{Centre national d'\'{e}tudes spatiales} 
 
\newacronym{FWHM}{FWHM}{Full Width at Half Maximum} 

\newacronym{NKG}{NKG}{Nishimura-Kamata-Greisen} 

% \end{acronym}



%\title{Enhancing the effective Fe abundance in UHECR sources}

\title{A setup for the precision measurement of multianode photomultiplier efficiency}

%\title{The Effect of a Maximum Source Energy Distribution on the Fe-to-proton ratio in Ultra-High-Energy Cosmic Rays}


\author[APC]{C. Blaksley\corref{cor1}}
\ead{blaksley@in2p3.fr}

\author[APC]{P. Gorodetzky}
\ead{philippe.gorodetzky@cern.ch}

% \author[LAL]{Others}
\cortext[cor1]{Corresponding author}

\address[APC]{Laboratoire Astroparticule et Cosmologie (APC), Universit\'e Paris 7/CNRS-IN2P3 UMR 7164, 10 rue A. Domon et L. Duquet, 75205 Paris Cedex 13, France}
% {Laboratoire Astroparticule et Cosmologie (APC), Universit\'e Paris 7, CNRS-IN2P3 UMR 7164, Paris, France}
% \address[LAL]{}

\begin{abstract}
In many applications, such as the detection of ultra-high energy cosmic rays using the air fluorescence method,
the number of  photons incident on the detector must be known. This requires a precise knowledge of the absolute efficiency of the photodetectors used. 
We present an experimental setup for measuring the single photoelectron gain and efficiency of multi-anode photomultipliers with a total uncertainty on
the order of a few percent. This precision is obtained by using a comparison to a NIST calibrated photodiode, and the presented method can be applied to both vacuum photomultiplier tubes and
other photodetectors.
This work is motivated by the need to calibrate the focal surface of the EUSO-Balloon instrument, which is a technical pathfinder for the future JEM-EUSO mission.
A complete discussion of photomultiplier calibration is presented and the efficiency measurement technique is discussed in detail. Example results are given to illustrate
the key points of the method. 
\end{abstract}


\end{frontmatter}

\section{Introduction}

In many applications knowledge of the absolute number of photons incident on a photodetector is needed with a reasonably low uncertainty. 
One example is the use of the air fluorescence method to detect \gls{EAS} created in the atmosphere by \glspl{UHECR} \cite{Greisen:1972zz, AFt3, AFt2}.
This method is employed both in current observatories, such as the Pierre Auger Observatory \cite{Abraham:2008ru, Abraham:2010mj} and Telescope Array \cite{AbuZayyad:2012ru},
and will also be used by future missions such as JEM-EUSO \cite{AdamsJr201376v2} 
and its pathfinder EUSO-Balloon \cite{Osteria:2013xua}. 
 
The air fluorescence method provides a calorimetric measurement of the energy deposited in the atmosphere by \gls{EAS}, 
but reconstructing the energy of the primary cosmic ray requires the absolute number of photons received by the detector.
At the same time, the number of photons emitted per eV of deposited energy, known as the air fluorescence yield \cite{Arqueros:2009zz, Rosado:2011qi}, must also be measured. 
A precise knowledge of both the photodetector efficiency and the atmospheric transmission are therefore essential to this observational technique. 

The photo-detection element in EUSO-Balloon is the Hamamatsu R11265-M64 \gls{MAPMT}.
The M64 is a \gls{PMT} with 64 individual anodes (pixels), each with an area of 2.88 mm$^{2}$, and an ultra bi-alkali photocathode with a quantum efficiency of 35-45\% for light in the 290 to 430 nm wavelength range. 
This MAPMT multiplies photoelectrons emitted from the photocathode using a stack of 12 metal-channel dynodes, giving a typical gain of $10^{6}$ at a cathode bias voltage of 900 V.

Each stage of a \gls{PMT} can be characterized by one or more parameters which, together, give the primary response characteristics of the PMT. 
The quantum efficiency $\epsilon_{q}$ characterizes the efficiency of converting photons into electrons at the photocathode.
The spectral sensitivity of the PMT is determined by the optical properties of the PMT entrance window and the variation of $\epsilon_{q}$ with wavelength.
As the photocathode is not perfectly uniform, the quantum efficiency depends on the location of the incident photons and
their angle of incidence. 

The collection efficiency $\epsilon_{\text{coll}}$ is the efficiency of collecting created photoelectrons into the multiplication stage of the \gls{PMT}, i.e.\ the probability that photoelectrons from the photocathode will land on the effective area of the first dynode.
The collection efficiency depends on the electrostatic field between the photocathode and the first dynode, as this affects the trajectories taken by photoelectrons.
Due to this, $\epsilon_{\text{coll}}$ depends on the applied voltage and hence the \gls{PMT} gain.

The single photoelectron efficiency is the ratio of the number of single photoelectron pulses on the \gls{PMT} anode to the number of photons incident on the cathode. 
It is the product of the quantum and collection efficiencies:
\begin{equation}
 \label{eq:DetectionEff}
  \epsilon(\lambda,x,y,V)= \epsilon_{q}(\lambda,x,y,\theta)\cdot\epsilon_{\text{coll}}(V, x,y)
\end{equation}
and is dependent on the wavelength $\lambda$, the incidence angle $\theta$, and location $(x,y)$ of the incident photons on the photocathode and the voltage $V$ between cathode and the first dynode.
The \emph{useful} efficiency also depends on the threshold used to separate single photoelectron pulses from noise.
% Both the single photoelectron gain and efficiency are highly sensitive to variations in the \gls{PMT} high-voltage power supply.
% , which include voltage drift and ripple, changes with temperature, input regulation fluctuations, and load regulation issues. 
%  \cite{HamamatsuManual3a}.

The single photoelectron gain $\mu$ is the number of electrons at the anode for each photoelectron collected, and is a function of the voltage on each element (dynode) of the PMT. 
There is a fluctuation in the number of electrons from shower to shower due to the nature of secondary emission. 

The single photoelectron characteristics of the PMT can be contrasted with the gain of the \gls{PMT}, which is the product of the single photoelectron
gain and the collection efficiency $\epsilon_{\text{coll}}$ of the first dynode. This property of the PMT is often measured as the ratio of the anode current to the cathode current, $G = I_{\text{a}}/I_{\text{k}}$. 
The current through the anode is the product of the single photoelectron gain, the efficiency, and the number of photons arriving at the photocathode per second 
(for currents which are too large, however, the anode response is no longer linear).

\subsection{Photomultiplier Calibration}
\label{sec:PMTCalibration}
 
The relative quantum efficiency of the cathode as a function of wavelength is generally well-characterized, but the absolute \gls{PMT} efficiency is more difficult to measure and is typically known with a precision of the order of 10--20\%.

% , as it is difficult to measure the collection efficiency independently. 
For many experiments, such as JEM-EUSO, this uncertainty is too large, and a method for measuring the absolute efficiency with an accuracy of a few percent is needed.  

In principle, $\epsilon_{q}$ can be translated into $\epsilon$ by measuring the collection efficiency using the ratio of the \gls{PMT} gain and the single photoelectron gain.
This method is difficult and imprecise for several reasons, however.
The cathode current $I_{\text{k}}$ must be at least several picoamperes to be measurable, which requires a very high light level, but at the same time the anode response is not linear at very high currents.
One way to overcome this is to operate the \gls{PMT} with only the cathode and first few dynodes polarized (so-called diode mode), so that $I_{\text{k}}$ can be measured at a very high light level without saturating the anode. 
After $I_{\text{k}}$ is measured, the light flux can be attenuated several orders of magnitude using neutral density filters with a known attenuation coefficient $\alpha$, and the 
\gls{PMT} is then put at full gain so that the anode current can be measured assuming a cathode current of $I_{\text{k}}/\alpha$. 
There can be large systematic uncertainties from the attenuation, and from not accounting for the dark current component of $I_{\text{k}}$.
Care must be also taken that there are no fluctuations in the emission of the source between the two measurements. 

A more precise technique is to measure the single photoelectron spectrum of the \gls{PMT}, i.e.\ the reponse at the anode to a single photoelectron.
The single photoelectron efficiency of the \gls{PMT} is the ratio of the number of detected signals, given by the surface of the one photoelectron peak, and the number of incident photons. 
In this case, the absolute number of photons incident on the photocathode must be determined.
This can be done by
\begin{inparaenum}[i\upshape)]
\item illuminating the \gls{PMT} with a calibrated source, or
\item comparing the \gls{PMT} to another calibrated detector. 
\end{inparaenum}

A calibrated light source can be provided by any source which gives a known number of photons per second per steradian. 
Examples include calibrated lamps, lasers, synchrotron radiation, Cherenkov emission, or other well known physical phenomena, such as Rayleigh scattering (e.g.~ref.~\cite{Kawana:2012cj}).  
If the power spectrum $dP^{3}/dS d\lambda d\Omega$ of the source is known with high accuracy, then the emissive surface, flux, and solid angle can be accounted for in the measurement.
This is a delicate task experimentally, and the variation of the flux with temperature and time must also be taken into account.

If the solid angle subtended by the emission of the source is small, such as for a laser, the flux is generally high in the emission region compared to the operational range of the \gls{PMT} and must be attenuated by several orders of magnitude. 
For sources in which the emission subtends a large solid angle, such as many calibrated lamps, the flux may require less attenuation, but then a precise knowledge of the spatial variation is critical.
The uncertainty involved in attenuating light sources in a controlled way can reach nearly 20\%, and so the precision on the measurement of the efficiency using this type of method is limited \cite{tubs091}. 
The uniformity of the source is also a problem in either case as attenuation filters can create lobes and other spatial variations in intensity which can be a source of systematic error. 

% Using Cherenkov or synchrotron radiation is subject to errors in calculating the Cherenkov yield, determining the effective aperture, and controlling the viewing distance. For radioactive Cherenkov sources there are also 
% systematic uncertainties due to photons trapped in the source by total internal reflection. There can also be an effective reduction in the number of photons incident on the \gls{PMT} due to photon coincidences when using Cherenkov emission \cite{Biller:1999ik}.
% The chance of photon coincidences is increased because the distribution of Cherenkov light from each charged particle emitted is focused in a cone around the direction of the charged particle.

% \subsection{Comparison to a Calibrated Detector}
Comparison to a calibrated detector, on the other hand, effectively creates a continuously calibrated source, and so eliminates the problem of intensity variations with temperature and time. However, the spatial variation of 
the source must still be taken into account.
Any absolutely calibrated detector can be used as the reference; here a photodiode is used.
A key point is that the same flux must be viewed simultaneously by both the \gls{PMT} and the photodiode, meaning that the gain of the \gls{PMT}, on the order of $10^{6}$ or higher, must be matched to the gain of the photodiode, which is slightly lower than one.

The calibration technique of Biller et al.~\cite{Biller:1999ik} attempted to overcome this by placing the reference photodiode closer to the source than the \gls{PMT}. 
In such a setup, the total flux at the \gls{PMT} must be inferred from the solid angle subtended by the \gls{PMT} and the measured flux per steradian. Both of these inferences are  potential sources of uncertainty.
Comparison to a calibrated detector thus represents a clear improvement over the use of a calibrated source, but it also presents difficulty due to the need to match the gain of the reference detector to that of the \gls{PMT} and in 
measuring the luminosity distribution of the source.

The setup of Biller et al.\ was further developed by Lefeuvre et al.~\cite{Lefeuvre:2007jq}, who applied the technique to a precise measurement of the air fluorescence yield.
This absolute calibration compares the response of the \gls{PMT} \emph{directly} to an absolutely calibrated photodiode.  
In this direct comparison method the \gls{PMT} to be calibrated and the reference photodiode view the same light in real time, and their gains are matched by attenuating the light by a factor of $\sim 10^{6}$--$10^{7}$ in a stable and repeatable way.
To do this, an integrating sphere is used as a stable and well-characterized splitter. The attenuation is itself measured using a second absolutely calibrated photodiode. This second photodiode directly replaces the \gls{PMT} so that it is at the same 
distance and uses the same opening of the sphere. In this way the intensity and spatial uniformity are the same in both measurements, resulting in an absolute measurement of the efficiency with a total uncertainty on the order of a few percent.

Here a calibration setup employing a developed version of the technique used by Lefeuvre et al.\ is presented, and, notably, the calibration procedure is fully explained. Special attention is paid to the characterization of
\gls{MAPMT}.

\section{Measurement Setup}
\label{sec:Measurement Setup}

\subsection{The Testing Environment}
\label{subsec:The Testing Enviroment: A Black Box}
The most essential element of the calibration setup is the measurement environment; the setup must be placed inside a black box to avoid exposing the high-gain \gls{PMT} to light. 
This allows a good signal-to-noise ratio to be obtained and avoids damaging the \gls{PMT} by over-exposure.
% As an example, say that we wish to test a \gls{PMT} with a 20 cm$^{2}$ photocathode at the single photoelectron level using a discriminator on the anode signal of the \gls{PMT}, giving a measured count rate. 
% We illuminate the \gls{PMT} with a constant light, so that photons arrive at the photocathode randomly and with some average rate. 
% If the double-pulse resolution of the discriminator is 40 ns, then a light level which gives a count rate of $\approx$ 25 kHz, would be at single photoelectron mode, i.e., there would be less than 1\% missed counts due to having two photoelectrons emitted in the same 40 ns.
% If the efficiency of the \gls{PMT} is 25\%, then a signal-to-background ratio of would 100 require that the flux of photons in the black box be less than 50 per second per cm$^{2}$, integrated in the entire spectral sensitivity of the \gls{PMT}. 
% The allowed background rate scales with the size of the \gls{PMT}, a single larger photocathode requires a lower background flux.

To eliminate as much background light as possible, the entry door to the black box is sealed using a camera baffle, which assures that the door is light tight each time it is closed. 
The box is constructed using lap joints, so that there are no small gaps at corners or between sides, which would give a straight path for photons to enter the box.
There should not be any holes in the box, and so every cable which goes through the black box must pass through a connector. Each connector should be light-tight. 
Inside the box, every surface is painted matte black and/or covered with black velvet to minimize reflections, and any light generated inside the box (e.g.\ on electronics) must be eliminated.

 \subsection{The Calibrated Source}
 \label{subsec:Integrating Spheres}

\begin{figure}[]
\centering
 \subfigure[Taking the PMT Spectrum]{\label{fig:CalibrationSetup:PMT} \includegraphics[angle=0,width=0.49\textwidth]{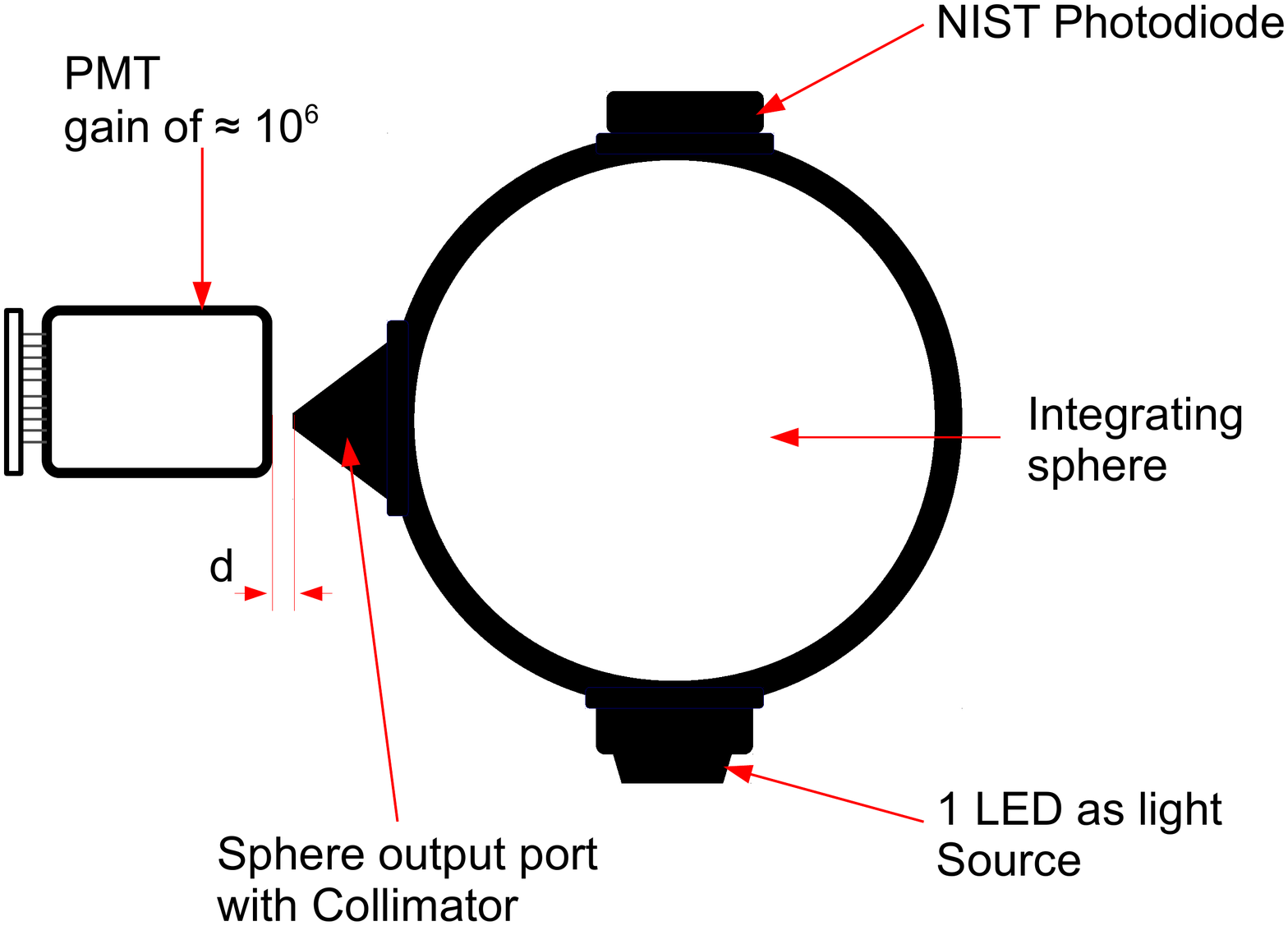}}
  \subfigure[Calibrating the Attenuation]{\label{fig:CalibrationSetup:NIST} \includegraphics[angle=0,width=0.49\textwidth]{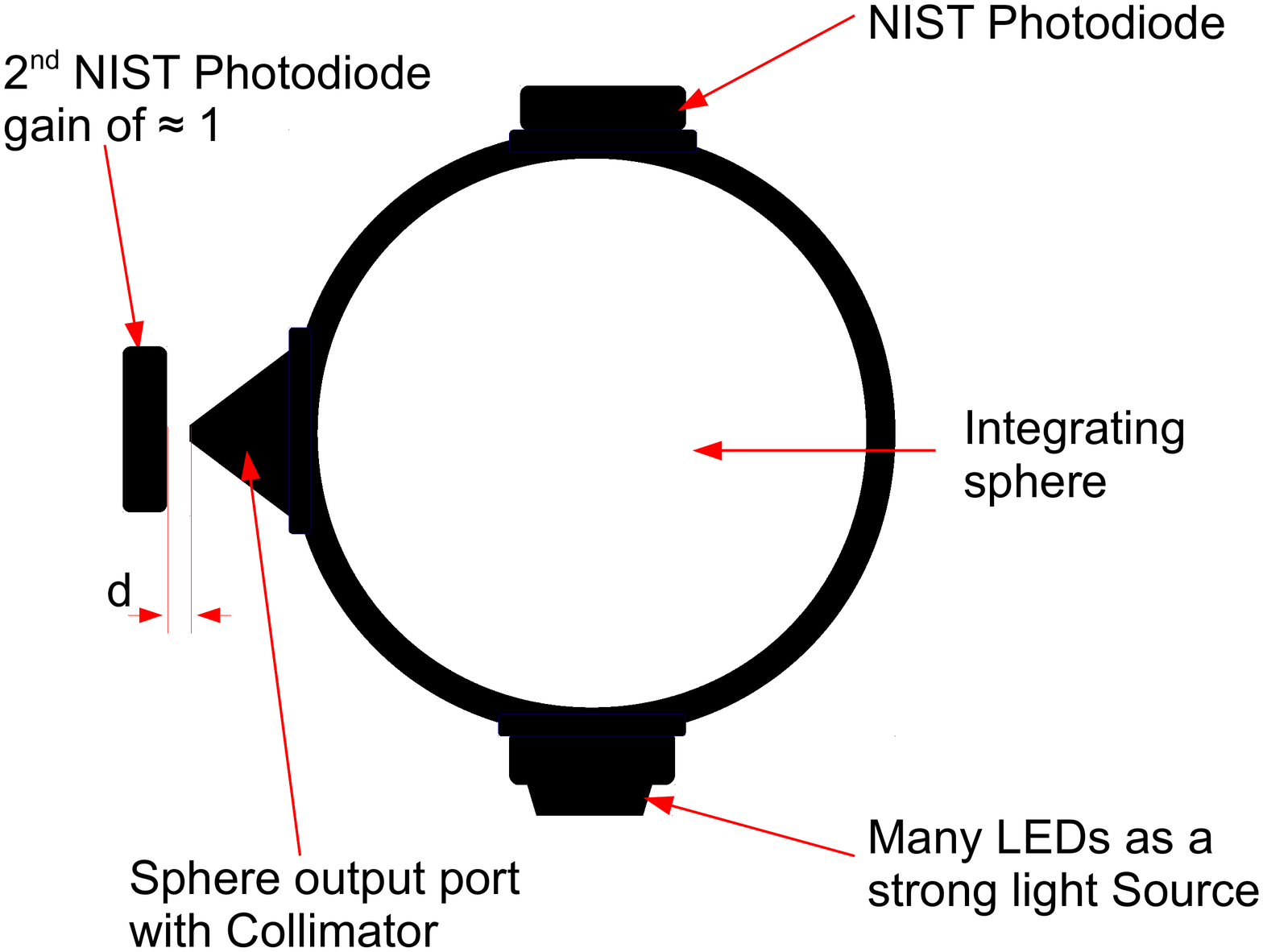}}
\caption[The Calibration Procedure]{ \label{fig:CalibrationSetup} Two diagrams showing the overall procedure for measuring the PMT efficiency. \fig\ref{fig:CalibrationSetup:PMT} shows the measurement of 
the single photoelectron spectrum of the PMT or MAPMT pixel. The attenuation of the integrating sphere and collimator assembly is calibrated in the second step, as shown in \fig\ref{fig:CalibrationSetup:NIST}.
}
 \end{figure}

The calibrated light source is provided by a monochromatic \gls{LED} combined with an integrating sphere and a reference detector. 
A diagram of the calibrated source is shown in Fig.~\ref{fig:CalibrationSetup}. 
An integrating sphere is a hollow sphere with a diffusive material coating the inside surface.

An integrating sphere diffuses radiant flux, and
the fraction of the radiance received by a finite interior surface of the sphere is proportional to the ratio of its area to the total interior surface area.
This property is independent of the viewing angle between the surfaces, the distance between them, and the size of the emitting part of the surface \cite{LabSphere02}.
The stability of the fraction of the flux received by each port of the integrating sphere is an important benefit compared to the sensitivity of a beam splitter. 
% For rapidly varying light signals, such as short pulses, the output signal can be noticeably distorted by multiple diffuse reflections. 
% This effect is not important in this application, however, because timing information it not required for the absolute calibration.

The integrating sphere used in this setup was manufactured by Labsphere.
%  and is shown in \fig\ref{pic:NISTfigures:ISpicture}. 
This integrating sphere has an internal diameter of 10.16 cm (4 in) and three ports, each located 90$^{0}$ from one another. The largest port has a diameter of 3.81 cm (1.50 in), and the two smaller ports each have 
a diameter of 2.54 cm (1.00 in). The interior of the sphere is coated with a proprietary Spectralon$^{\text{\textregistered}}$ material which has a diffuse reflectivity of 
$\approx0.97$ for 400 nm light. The Spectralon coating is a purified PTFE with a thickness of 8 mm, and, due to this thickness, it is difficult to calculate the effective area of a port precisely.
We therefore did not perform any simulations or calculations using the sphere parameters, but instead measure the emission intensity from the sphere output port.  

% \begin{figure}[]
% \centering
% %  \subfigure[The Integrating Sphere]{\label{pic:NISTfigures:ISpicture} 
% \includegraphics[angle=270,width=0.45\textwidth]{IntegratingSpherePicture_rotated}
% % }
% %  \subfigure[A Diagram of the NIST Photodiode]{\label{fig:NISTfigures:diagram} \includegraphics[angle=270,width=0.45\textwidth]{NISTdiagram}}
% \caption[Picture of the Integrating Sphere Assembly]{ \label{fig:NISTfigures} 
% % \fig\ref{fig:NISTfigures:diagram} shows a diagram of the NIST photodiode with dimensions, taken from the Ophir data sheet \cite{OphirDataSheet}. 
% % As can be seen, the sensitive area is about approximately 11 mm$^{2}$. \fig\ref{fig:NISTfigures} shows
% A photograph of the integrating sphere, as it is setup inside the black box.
% On the top port, directly facing the camera, is the NIST photodiode. 
% The LED is mounted on the lower port. The output port is on the left, with an attached collimator. 
% The sphere itself is mounted on an X-Y stage, allowing a precise placement of the light spot on the \gls{PMT} photocathode.
% }
% \end{figure}

A silicon photodiode (PD300-UV by Ophir \cite{OphirDataSheet}) was used as the reference detector for this setup. 
Silicon photodiodes can be operated unbiased, and in this configuration they are stable over almost 10 decades of incident power.
The area of this photodiode is about 120 mm$^{2}$, and it is sensitive to the wavelength range from 200 nm to 1100 nm.
The operational power range of the photodiode is from 20 pW to 3 mW with a resolution of 0.001 nW, and the output noise level (RMS) is on the order of $\pm$ 1 pW. 
The temperature dependance of the photodiode response is less than 0.1\%  per $^{\circ}$C for light of wavelength less than 800 nm.

The most important characteristic of this photodiode is its wavelength-dependent absolute efficiency.
The calibration of this efficiency is given by the \gls{NIST} with an uncertainty ($\sigma$) of 1.5\% in the wavelength range $270$--$950$ nm.

In the present setup the photocurrent generated in the NIST photodiode is measured using a specially designed ``LaserStar'' power meter, which is essentially a pico-ammeter which uses the absolute calibration curve of the photodiode.
Once the wavelength of the source is selected on the ammeter it gives a direct reading of the power incident on the photodiode, with a precision on the measured photo-current of 0.5\% \cite{LaserStar} 
(to be added quadratically to the 1.5\% uncertainty of the NIST calibration).
 
\subsection{Gain Matching} 
\label{subsec:Absolute Calibration Setup}

\begin{figure}[]
\centering
\includegraphics[angle=0,width=0.49\textwidth]{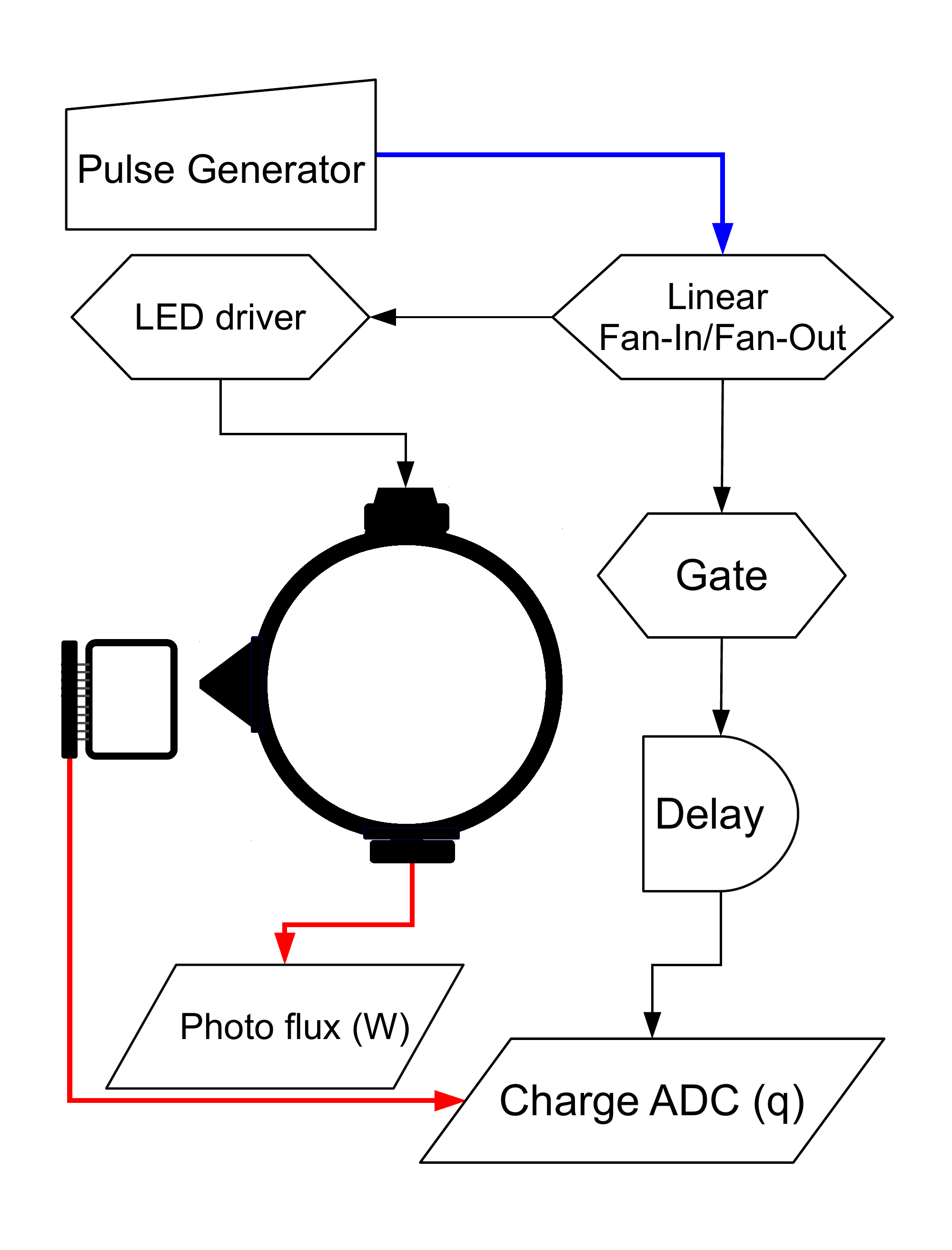} 
\caption[Signal Logic of the Calibration Setup]{ \label{fig:CalibrationSetup:Logic} The signal logic of the calibration setup (e.g., implemented with NIM modules). 
% This diagram is to illustrate the method; details for our actual setup, are omitted for clarity. 
From the top of the figure:  A short pulse of the desired rate is created using a pulse generator and is sent to
a fan-in/fan-out logic module. The resulting pulses drive an LED and generate the gate which is used as the integration window for a QDC. 
}
 \end{figure}

In order to match the gain of the NIST photodiode to that of the \gls{PMT} it is necessary to determine the attenuation required. 
The light is provided by one or more monochromatic \glspl{LED}, which are placed on the larger, 3.81 cm, port of the integrating sphere.
The LED(s) are pulsed in coincidence with the integration windows (gates) of the charge measurement (see~\fig\ref{fig:CalibrationSetup:Logic} and sect.~\ref{subsec:PMT readout electronics}).

The photodiode is placed on one of the 2 smaller integrating sphere ports. The aperture to the photodiode was 9 mm to limit edge effects from the material surrounding it, and
the \gls{PMT} was illuminated through the 2nd small port. 
% , with an added diaphragm to reduce the effective size of the port, and thus the flux the \gls{PMT} receives.
The $\pm1~$pW noise level of the photodiode can be used to estimate the minimum background. From this, there must be at least 0.1 nW on the photodiode to have a good signal-to-noise ratio during the reference measurement. 
If the light source is a monochromatic LED with a wavelength of 398 nm, and the conversion efficiency of the photodiode\footnote{The PD300-UV silicon photodiode gives a photocurrent of 148 mA/W at 398 nm.} is about 25\%, 
this would correspond to $\approx 8~10^{8}$ photons per second incident on the surface of the photodiode. 
% The reduction in flux between the \gls{PMT} and the photodiode must be $\sim10^{7}$ or more to match the gain of the two.

From the Poisson statistics, the number of gates containing two photoelectrons in the spectrum is less than 1\% of the number of one photoelectron gates when $\sim99\%$ of all gates give no photoelectron. 
If the LED is pulsed at a rate of $\sim1$ kHz and the efficiency of the \gls{PMT} is around 25\%, this corresponds to no more than $40$ photons per second incident on the \gls{PMT}. 
The reduction in flux between the \gls{PMT} and the photodiode must then be $\sim10^{7}$ to have both 40 photons per second on the \gls{PMT} and a good measurement of the incident power using the photodiode. 
% If the LED pulse rate is increased, however, so too is the integrated power on the photodiode. This means that the power delivered to the photodiode per pulse can be lower with the total energy delivered to the photodiode remaining the same.
% The total attenuation needed can therefore be reduced by increasing the overall pulse rate of the LED (and thus the data acquisition rate). 

An attenuation of $\sim10^{7}$ could be given by a single integrating sphere by using a diaphragm with a radius on the order of $10^{-4}$ cm on the output port of the sphere. 
One alternative solution is to add a second integrating sphere, as done by Lefeuvre et al.~\cite{Lefeuvre:2007jq}. This method is patented \cite{GoroCaibrationPatent}, but is not ideal, as an integrating sphere is a Lambertian source. 
If the PMT is instead placed some distance from the output port of the (first or second) integrating sphere, then the illumination will be largely uniform, 
but the sensitive surface of both the photocathode and of the photodiode must be taken into account.

Another possibility is to use a collimator at the output port of the integrating sphere in order to illuminate the \gls{PMT} in a more uniform way, and simultaneously restrict this illumination to a small area of the photocathode. 
In this case the incidence angle of the incoming photons will also be well-controlled. The exact dimensions of the collimator are based on the needed spot size and the required attenuation. 
The attenuation of the collimator is not calculated, but rather is measured directly in the second step of the calibration procedure.

An important consideration is that a parallel beam of light, such as from a collimator, will give a measurement of the efficiency over an area of the photocathode equal to the spot size of the beam. 
A uniform illumination of the entire photocathode, for example, will give the efficiency averaged over the entire photocathode of the \gls{PMT}. 
Illumination of the entire cathode is thus more appropriate for relative efficiency measurements -- i.e.\ comparing one PMT or one MAPMT pixel to another, while a collimated beam with a small enough spot size that the measured area of the PMT photocathode
can be replaced with a photodiode is more suitable for measuring the absolute efficiency.

%  A central ``clock'' pulse is created with a pulse generator and sent to a linear fan-in/fan-out circuit which outputs two copies of the clock pulse. One copy of this pulse is sent to a LED driver. The other 
% copy is used to create the gate for the anode charge measurement. This can be done using a discriminator with an adjustable width, for example.
% The gate pulse can be set in coincidence with a timer to control the time of the run. 
% The gate is delayed relative to the LED pulse to account for the rise time of the LED, the transit time of the cables connecting the PMT anode to the read-out, etc.

%--------------------------------------------------------------
% \subsection{PMT sorting Setup}

\subsection{Read-out electronics}
\label{subsec:PMT readout electronics}

\glspl{PMT} are essentially current sources, with each single photoelectron collected resulting in a shower of electrons at the anode. 
The single photoelectron response of a \gls{PMT} is thus best measured as a charge spectrum. 
This is achieved using \gls{QDC} electronics, which measure the total charge received during a defined time window, known as a gate.
A diagram the setup is shown in \fig\ref{fig:CalibrationSetup:Logic}. 
The \gls{LED} is pulsed in coincidence with the integration gate of the \gls{QDC}.

For \gls{MAPMT} this is complicated by the need to read out multiple anodes from each PMT. In the case of the M64 \gls{MAPMT}, sixty-four simultaneous read-out channels are required (in order to test one PMT at a time).
As the gain of the M64 is of the order of $10^{6}$, a single photoelectron pulse gives around 160 fC at the anode. 
\glspl{QDC} with a charge resolution on the order of 20 fC or sixty-four high performance integrating amplifiers are therefore required in order to take 
clean single photoelectron spectra.

\begin{figure*}[p]
  \centering
  \includegraphics[width=1.0\textwidth]{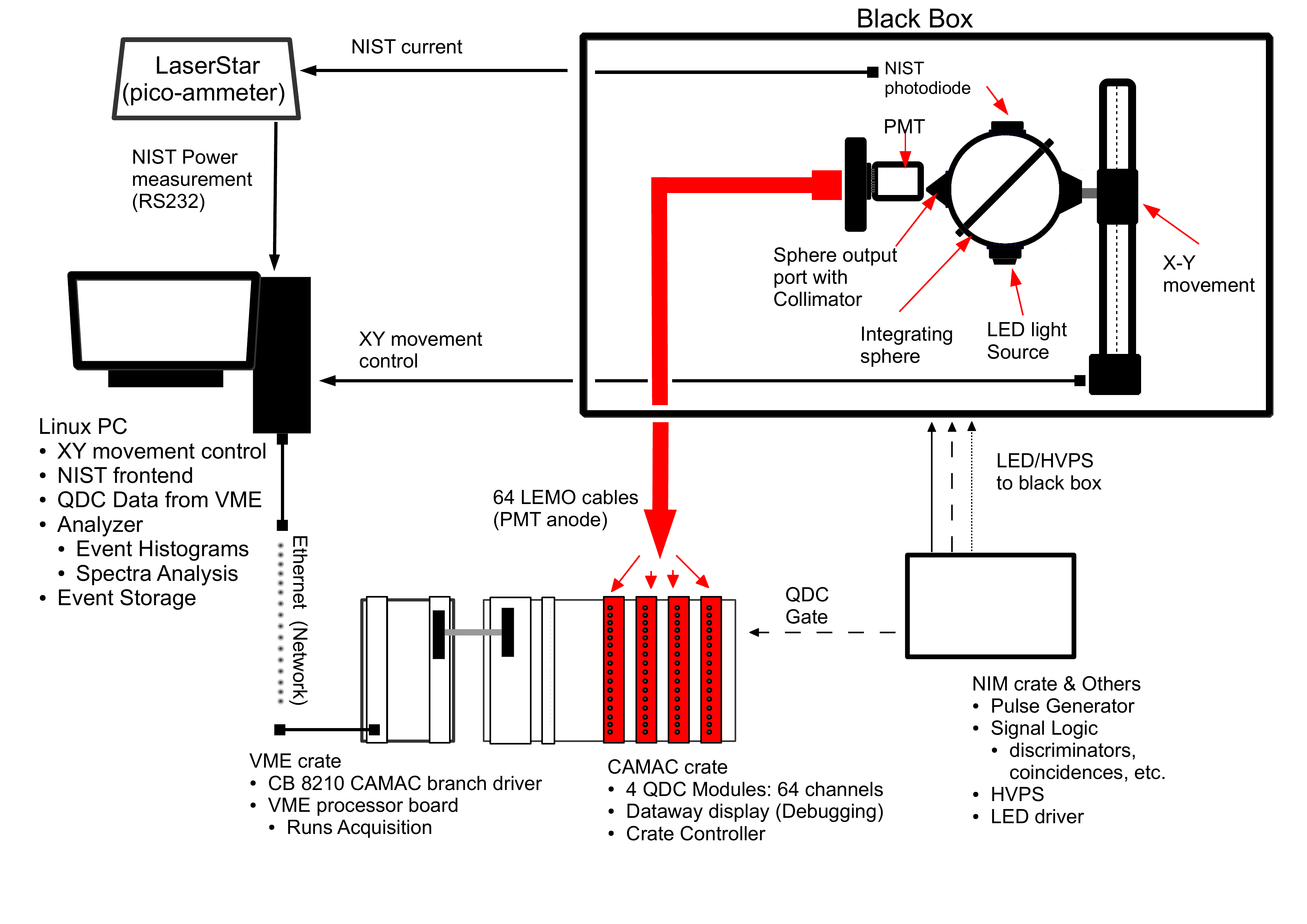}
  \caption[The PMT Sorting Setup]{\label{fig:DAQdiagram} A diagram of the \gls{PMT} calibration setup. 
The \gls{PMT} is placed inside a black-box and is illuminated using the output of an integrating sphere, as described in sect.~\ref{sec:Measurement Setup}. }
 \end{figure*}

A diagram of the \gls{DAQ} is shown in \fig\ref{fig:DAQdiagram}.
The core of this system is a standard CAMAC crate containing 4 CAEN C1205 charge-to-digital conversion modules.
These modules have a charge resolution of 20 fC per channel or better.
The CAMAC crate is controlled by a SEN CC 2089 ``A2'' crate controller, and the CAMAC branch is driven by a CBD 8210 branch driver, which interfaces the CAMAC branch to a VME crate. 
The VME crate is controlled by a Motorola MVME 3100 VME processor board running Debian Linux on 32-bit Power PC architecture.

A data acquisition program was written for this setup in C/C$++$ on a Linux PC using the \gls{MIDAS} framework \cite{MIDASv2}.
\gls{MIDAS} is a distributed data acquisition framework which allows complicated data acquisition and analysis routines to be implemented.
Here the program which controlled the VME and CAMAC crates ran directly on the VME processor board, which was connected to the backend PC by Ethernet.
The acquisition with setup was able to proceed at a rate of 2 kHz for 64 QDC channels in parallel, the rate being limited by the CAMAC signal definitions. 
As the C1205 is compatible with the FAST-CAMAC standard, an upgrade of the CAMAC crate controller to a FAST-CAMAC compatible model would increase the read-out rate by a factor of between 2 to 10 \cite{FASTCAMACb,FASTCAMACa}.

\label{sec:TheDAQ}

In addition to the program driving the \glspl{QDC}, the other hardware was also integrated into the acquisition software. This included two Zaber T-LSM200
movement supports mounted together, orthogonal to each other. These supports held the integrating sphere (light source), allowing the incident light spot to be positioned on the photocathode of the \gls{PMT} under test 
with an accuracy of 0.003 mm. The LaserStar power meter which reads the NIST photodiode was also interfaced with the DAQ. The photodiode was sampled at a rate of 10 Hz and this information was
stored event by event along with the QDC data and the X-Y position of light spot.

The \gls{DAQ} performed a full analysis of gathered single photoelectron spectra at the end of each measurement run using ROOT. This analysis included data gathered from the NIST photodiode, and
the results from the single photoelectron spectrum analysis were also interfaced with the control of the X-Y movement hardware, to allow, for example, centering with high precision on a given pixel using the 
response of the PMT itself. The extraction of measurements from the single photoelectron spectra will be discussed in sect.~\ref{subsec:The Single Photoelectron Spectrum}.

%--------------------
\section{Measurement Procedure}
\label{sec:Measurement Procedure}

\subsection{The Single Photoelectron Spectrum}
\label{subsec:The Single Photoelectron Spectrum}

\begin{figure*}[ht]
\centering
\includegraphics[angle=0,width=0.9\textwidth]{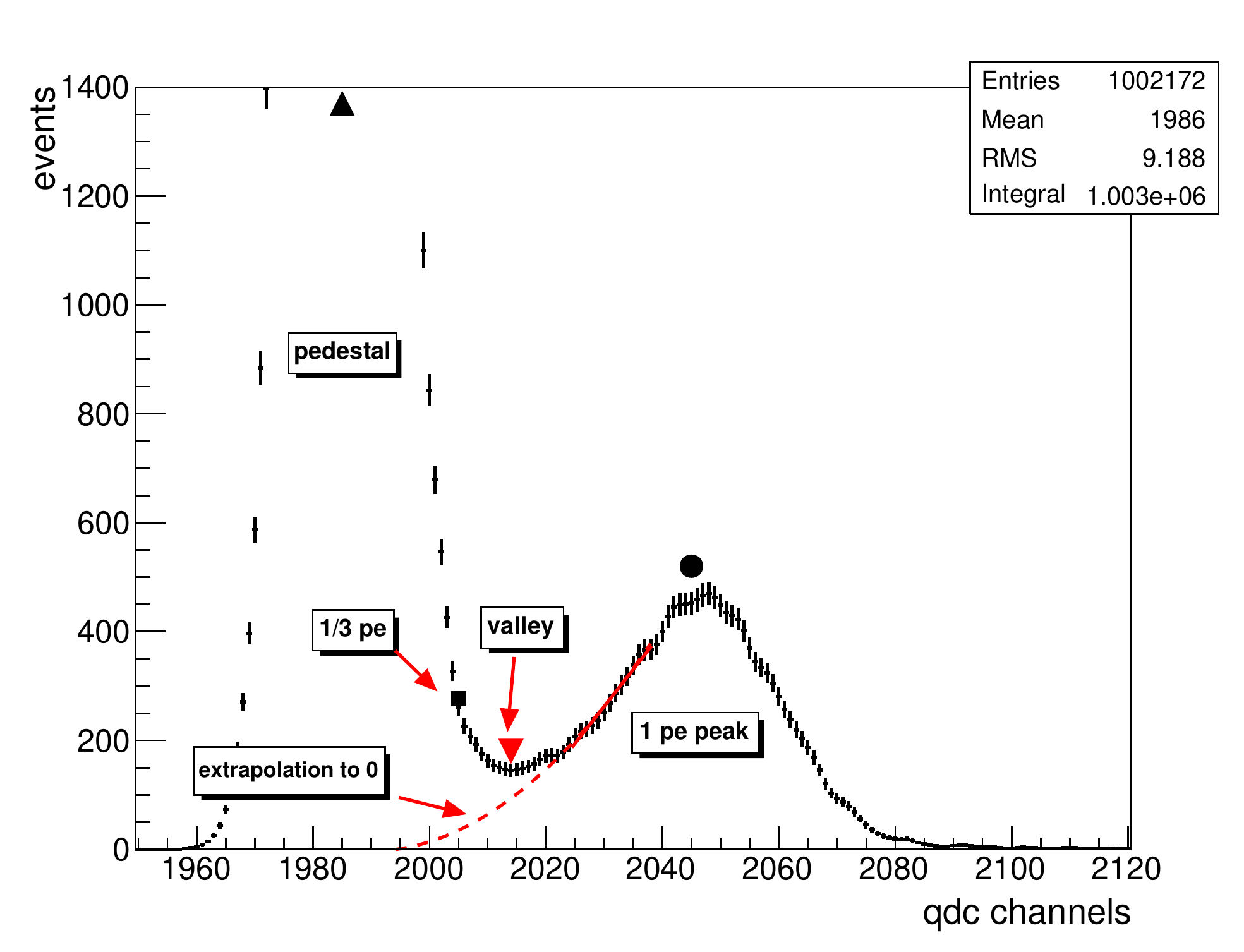}
\caption[Example of a Single Photoelectron Spectrum]{ \label{fig:ExampleSPEspectra} 
An example of a single photoelectron spectrum measured using the setup
presented in this article. The mean of the zero photoelectron (upward-facing triangle) and one photoelectron peaks (circle) are marked, as are the valley (downward-facing triangle) and 1/3 of a photoelectron (square).

}    
\end{figure*}

To operate in single photoelectron counting mode the probability that more than one photoelectron is collected within the time-resolution of the measurement must be negligible. 
Here the time-resolution is the QDC integration time, i.e.\ the gate width.
The average number of photoelectrons per gate is given by the product of the average number of photons arriving at the photocathode during each gate, the photocathode quantum efficiency, and the collection efficiency: $n_{\text{pe}} = n_{\gamma}\epsilon$.

The creation and collection of any given photoelectron is an independent process and can be assumed to occur at some constant average rate.
The probability of photoelectrons entering the electron multiplier is thus described by the Poisson distribution:
\begin{equation}
\label{eq:poisson}
p(n;\eta)= \frac{\eta^{n}}{n!}e^{-\eta}
\end{equation}
which gives the probability to find a given number of occurrences $n$, if they are independent and happen at an average rate per unit time $\eta$.

From \eq\ref{eq:poisson}), the ratio of the number of gates containing two photoelectrons to those containing a single photoelectron can be expressed in terms 
of the ratio of gates containing zero photoelectrons to those containing one photoelectron as
\begin{equation}
\label{eq:SPEcondition}
\frac{p(2;\eta)}{p(1;\eta)} = \frac{\eta^{2}}{2!}\frac{1!}{\eta} = \frac{\eta}{2} = \frac{1}{2}\frac{p(1;\eta)}{p(0;\eta)}
\end{equation}
This allows the contamination of two photoelectrons in the single photon electron spectrum to be estimated: if the ratio of one photoelectron counts to pedestal counts is less than 1\%, then the ratio of two photoelectron counts
to one photoelectron counts is less than 0.5\%.

An example of a single photoelectron spectrum taken with the setup presented in this paper is shown in \fig\ref{fig:ExampleSPEspectra}. 
On the left hand side of the spectrum is the region of the spectrum, known as the pedestal, corresponding to gates during which no photoelectrons were collected into the electron
multiplier of the \gls{PMT}. In theory, the pedestal will be centered on zero charge, but in practice the pedestal has a non-zero mean due to leakage current from the PMT and the current offset of the read-out electronics.
The area of the pedestal gives the number of gates in which no photoelectrons were collected.

The peak on the right is the single photoelectron peak. Several key pieces of information about the \gls{PMT} can be extracted from the spectrum:
\begin{itemize}
  \item The mean charge of the single photoelectron peak minus the mean charge of the pedestal gives the single photoelectron gain $\mu$ in coulombs (after calibration of the \gls{QDC}).
  The width of the single photoelectron peak is due to the inherent fluctuation in the secondary emission at the first dynode, convolved with the width of the pedestal.
  \item The peak-to-valley ratio can be used as a figure of merit for both the spectrum and the \gls{PMT} itself.
  The pulse-height resolution of the \gls{PMT}, defined as the ratio of the \acrshort{FWHM} of the single photoelectron peak to the height of the peak, increases with peak-to-valley ratio. 
  \item The surface area of the single photoelectron peak gives the number of one photoelectron gates, i.e.\ the total number of single photoelectron counts. 
\end{itemize}
If the single photoelectron gain and peak-to-valley ratio are high enough, then it may be possible to resolve a third peak corresponding to two photoelectrons at twice the charge of the single photoelectron peak. 

% In addition to measuring single photoelectron spectra of \gls{PMT}s by taking a charge or pulse height spectrum, the equivalent information can be obtained in a counting experiment. To do this the anode signal of the \gls{PMT}
% is sent through an integrating preamplifier and then through a discriminator circuit. The discriminator gives a output pulse whenever it receives an input pulse with a voltage over a set threshold. 
% The plot of the count rate versus threshold is known as an S-curve.
% As the single photoelectron spectrum is the charge distribution of the \gls{PMT} signal, the S-curve is simply its cumulative distribution
% function (in volts, rather than charge, because of the integrating preamplifier), and so the single photoelectron spectrum can be recovered by derivation.

The absolute measurement of the detection efficiency of the \gls{PMT} proceeds as follows: 
\begin{inparaenum}[ 1\upshape)]
\item The first measurement is shown in \fig\ref{fig:CalibrationSetup:PMT}.
The \gls{PMT} is illuminated using the calibrated source described in sect.~\ref{subsec:Integrating Spheres} (a pulsed LED with an integrating sphere and collimator). 
A single photoelectron spectrum is taken using a \gls{QDC}, which integrates the 
total charge received during a gate generated in coincidence with the LED pulse. Both the gate length and delay are adjusted so that the anode pulses from the \gls{PMT} are contained within the gate.
The use of a gate--pulse coincidence reduces the contribution of dark pulses (i.e.\ real photoelectron pulses due to thermionic emission, etc.) to a negligible level.
 
The light level is then reduced until the number of gates which give one photoelectron is about 1\% of the number of gates which give no photoelectron.
From the Poisson statistics, the contamination of two photoelectrons in the spectrum is then around 0.5\% of the number of single photoelectrons.  

A single photoelectron spectrum is then taken with enough total gates to give the needed statistical uncertainty on the number of one photoelectron counts $N_{\text{pe}}$. 
At a rate of one photoelectron per 100 gates, this means one million gates are needed to reach a
$\mathcal{O}\left(1\%\right)$ statistical uncertainty. The total time $\tau$ over which the spectrum is taken is also measured, for example by putting a control timer on the pulse generator which gives the gate signal. 
The power $P$ received by the NIST photodiode attached to the integrating sphere is recorded simultaneously.

The resulting single photoelectron spectrum is analyzed to determine the number of single photoelectrons $N_{\text{pe}}$, which is threshold dependent. 
% $N_{\text{pe}}$ should therefore be determined for the actual working threshold.
$N_{\text{pe}}$ should be determined either for a chosen working threshold (for example 1/3 of the mean single photoelectron charge) or extrapolated to the zero of the pedestal.
It must be noted that in photon counting experiments the threshold is chosen so that it lies in the valley between the pedestal and single photoelectron peak. 

\item The \gls{PMT} is then replaced by a second NIST photodiode, as shown in \fig\ref{fig:CalibrationSetup:NIST}. The photodiode is placed at the same distance from the output of the collimator as the \gls{PMT} was previously.
The single LED is also replaced with a collection of LEDs, so that the power on the second photodiode is high enough to have a good signal-to-noise ratio.
Here the fact that the photodiode is linear across 10 decades of power is exploited.
The ratio of the power measured by the two photodiodes gives the attenuation:
\begin{equation}
 \label{eq:AttenuationFactor}
 \alpha = \frac{P_{\text{PMT}}}{P_{\text{sphere}}}
\end{equation}
where $P_{\text{sphere}}$ is the power measured by the photodiode on the sphere and $P_{\text{PMT}}$ is the power measured by the photodiode replacing the \gls{PMT}.
\end{inparaenum}

\section{Results}
Each single photoelectron spectrum is analyzed to extract the number of single photoelectron counts and the position of the single photoelectron peak.
In order to work in an efficient way with a large number of \glspl{MAPMT} this analysis was automated and built into the \gls{DAQ}.
The example spectrum shown in \fig\ref{fig:ExampleSPEspectra} was analyzed by this routine. 
% An additional example of sixty-four spectra, one for each pixel of a M64 \gls{MAPMT}, gathered in one run is shown in \fig\ref{fig:64SpectraAnalyzed}.

Each spectrum is first smoothed using the 353QH algorithm\footnote{The 353QH algorithm consists of taking a
running median of three, followed by a running median of five and another running median of three. The data is then quadratically interpolated, and the end result is
subjected to a running average.} \cite{Friedman:1974vj} to reduce statistical fluctuations, after which the valley, shown by the downwards-facing triangle in \fig\ref{fig:ExampleSPEspectra}, is found using a simple peak searching routine.
No attempt was made to perform any fits in the analysis as neither the single photoelectron peak, nor the 
pedestal, are properly described by a simple Gaussian distribution. A discussion of the deconvolution of PMT spectra is 
given in \cite{Bellamy1994468}, but such multi-parameter analyses are difficult, model dependent, and not needed in this case.

Once the valley is found, the mean of the photoelectron peak is determined, as shown by the circle in \fig\ref{fig:ExampleSPEspectra}. 
The dashed line in the spectrum is an extrapolation of the single photoelectron peak from the valley to the mean of the pedestal. This extrapolation is
used to provide an estimate of the physical efficiency independent of the choice of threshold.

\subsection{Single photoelectron gain}
The gain and efficiency are calculated from the results of the spectrum analysis. 
The single photoelectron gain $\mu$ is given by
\begin{equation}
 \mu = \frac{1}{e}(q_{1} - q_{0})
\end{equation}
where $q_{1}$ and $q_{0}$ are the mean charge of the single photoelectron peak and the pedestal respectively, and $e$ is the elementary charge.
This calculation includes the response of the \gls{QDC} (i.e., the conversion from QDC channels to absolute charge).
As the statistical uncertainty on the gain at a given supply voltage is negligible,
the uncertainty on the resulting gain is limited by the characterization of the \gls{QDC}, here performed individually for each of the sixty-four \gls{QDC} channels with an uncertainty of 1\%. 

\subsection{Single photoelectron efficiency}
\label{subsec:Single photoelectron efficiency}
The efficiency $\epsilon$ of the \gls{PMT} is given by the equation
\begin{equation}
 \label{eq:EfficiencyCalculation}
 \epsilon = \frac{N_{\text{pe}}}{N_{\text{photons}}} = \frac{N_{\text{pe}}}{P\alpha\tau}\frac{hc}{\lambda}
\end{equation}
 where
% \begin{itemize}
%  \item $N_{\text{pe}}$ is the number of single photoelectron counts in the spectrum
%  \item $P$ is the average photodiode power measured while taking the spectrum
%  \item $\tau$ is the time of the measurement
%  \item $h$ is the Planck constant
%  \item $c$ is the speed of light in vacuum
%  \item $\alpha$ is the measured attenuated factor
% \end{itemize}
$N_{\text{pe}}$ is the number of single photoelectron counts in the spectrum,
$N_{\text{photons}}$ is the number of incident photons,
  $P$ is the average photodiode power measured while taking the spectrum,
 $\tau$ is the length of time of the measurement,
  $h$ is the Planck constant,
$c$ is the speed of light in vacuum, and
 $\alpha$ is the measured attenuation.

% \subsubsection{Uncertainty estimate}
The overall uncertainty on the measured efficiency can be separated into a systematic and statistical part:
\begin{equation}
 \label{eq:EfficiencyStatPlusSys}
 \left|\frac{\delta \epsilon}{\epsilon}\right| \leq \left|\frac{\delta \epsilon}{\epsilon}\right|_{\text{stat}} + \left|\frac{\delta \epsilon}{\epsilon}\right|_{\text{sys}}
\end{equation}
The systematic uncertainty is dominated by the uncertainty on the power measured by the reference photodiode. 
There is a further contribution, however, from the systematic underestimation of $N_{\text{pe}}$ due to the contamination of two photoelectrons
in the spectrum. This is less than $0.5\%$ if the single photoelectron spectrum is taken under the proper conditions ($\geq99\%$ of all events in the pedestal).
The same NIST photodiode is used in the measurement of the incident power while taking the spectrum and in the measurement of the reference power during the attenuation calibration. 
The systematic uncertainties on $P$ and $P_{\text{sphere}}$ are thus completely correlated and will cancel out in the quantity $P\alpha$. 
The systematic error is thus given by 
\begin{equation}
 \left(\frac{\delta \epsilon}{\epsilon}\right)^{2}_{\text{sys}}= \left(\frac{\delta P_{\text{PMT}}}{P_{\text{PMT}}}\right)^{2}+ \left(\leq0.5\%\right)^{2}
\end{equation}
The two sources of systematic error have been added in quadrature as they are independent.

The statistical uncertainty is dominated by the statistical uncertainty on $N_{\text{pe}}$, which can easily be brought below $\mathcal{O}\left(1\%\right)$.
The uncertainty on the Planck constant and $\lambda$ are completely negligible. The error on $\tau$ is also negligible in this case, but depending on the experimental setup -- i.e., the measurement rate, this may not necessarily be true.
The statistical uncertainty on the measurement of the attenuation is also negligible, as the mean value can be determined to arbitrary statistical precision, but
the uncertainty on the measurement of the mean power while taking the spectrum may not be negligible, depending on the sampling rate and the width of the sample distribution.
There is also a contribution to the statistical uncertainty on $P_{\text{0}}$ from the read out of the current from the photodiode. For the LaserStar pico-ammeter this is 0.5\%.

Each of these uncertainties are independent and random in nature and they can be added in quadrature:
\begin{equation}
 \left(\frac{\delta \epsilon}{\epsilon}\right)^{2}_{\text{stat}}= \left(\frac{\delta P}{P}\right)_{\text{stat}}^{2}+ \left(\frac{1}{\sqrt{N_{\text{pe}}}}\right)^{2}
\end{equation}
% From this the total error on the \gls{PMT} efficiency is:
% \begin{equation}
%  \label{eq:TotalGeneral}
%  \left|\frac{\delta \epsilon}{\epsilon}\right| \leq \left|\left(\frac{\delta P_{\text{0}}}{P_{\text{0}}}\right)_{\text{stat}}^{2}+ \left(\frac{1}{\sqrt{N_{\text{pe}}}}\right)^{2}\right|^{1/2}_{\text{stat}}
%  + \left| \left(\frac{\delta P_{\text{PMT}}}{P_{\text{PMT}}}\right)^{2}+ \left(\sim0.5\%\right)^{2}\right|^{1/2}_{\text{sys}}
% \end{equation}
Using the systematic error of the NIST calibrated PD300-UV photodiode (1.5\%),
taking a single photoelectron spectrum such that the statistical error on $N_{\text{pe}}$ is negligible and the number of two photoelectrons is 0.5\% the number of one photoelectrons, 
and assuming that the statistical error of $P$ is that of the LaserStar current read-out (0.5\%),
the total uncertainty on the efficiency using this technique is then
\begin{equation}
 \label{eq:Total}
 \left|\frac{\delta \epsilon}{\epsilon}\right|
 \leq \left(0.5\%\right)_{\text{stat}} + \left(1.6\%\right)_{\text{sys}}
\end{equation}
These systematic and statistical errors are independent and their sum in quadrature gives an estimate for the total uncertainty of
\begin{equation}
 \frac{\delta \epsilon}{\epsilon} = \pm~1.7\%
\end{equation}

\subsection{Measurement Example and discussion}
\label{sec:Example Measurement and discussion}

\begin{figure}[t]
  \centering
  \includegraphics[width=0.45\textwidth]{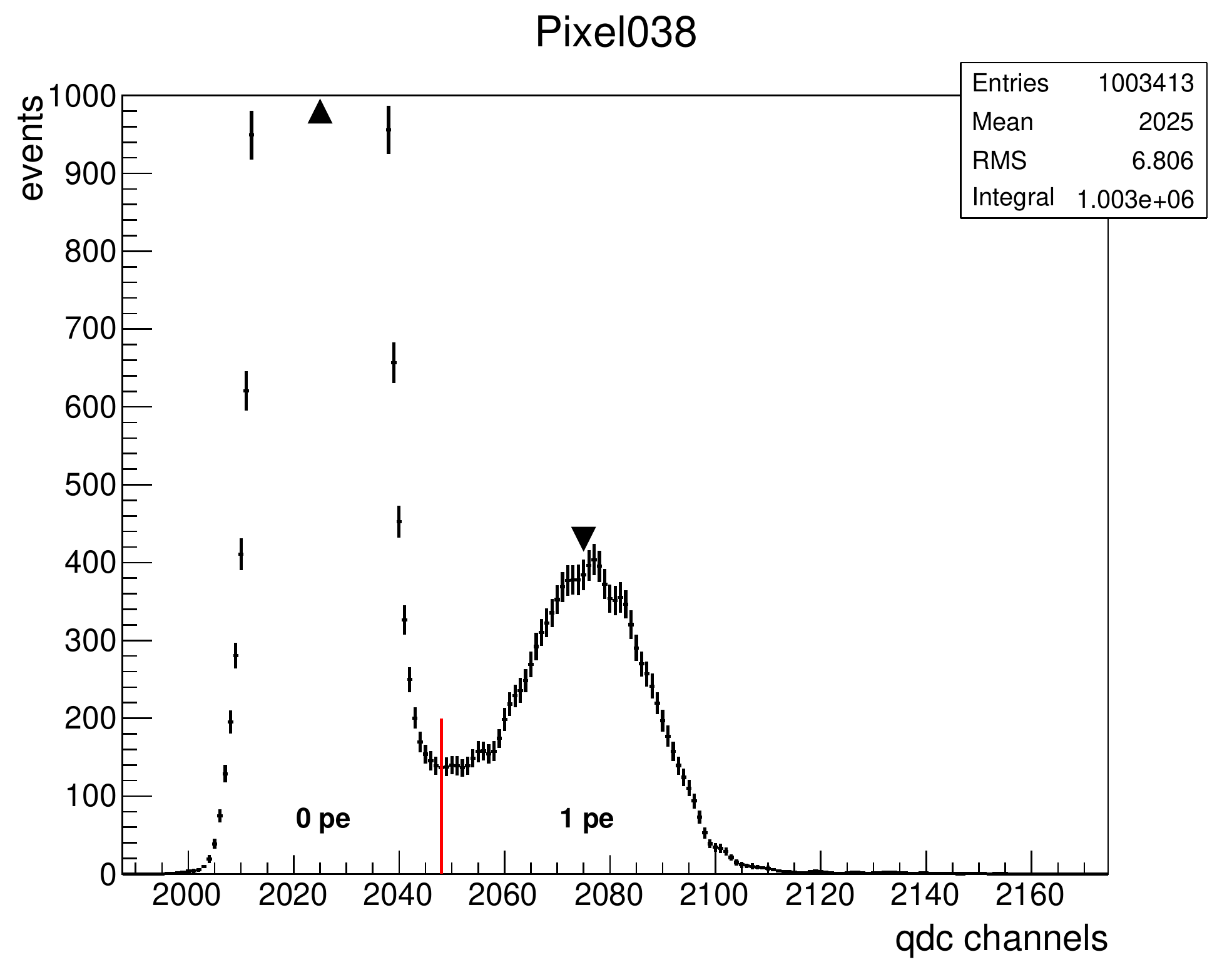}
  \caption[Example absolute spectrum]{A single photoelectron spectrum taken for one pixel of a Hamamatsu R11265-M64 \gls{MAPMT}. This spectrum was taken using the setup described in sect.~\ref{sec:Measurement Setup}.
The downward-facing triangle indicates the mean of the single photoelectron peak, while the upward-facing triangle marks the mean of the pedestal. The threshold, shown by the horizontal line, is set in the valley between the two peaks.
}
  \label{fig:Spectrum_EC107-PMTC-Pixel038}
 \end{figure}

A single photoelectron spectrum from one pixel of a Hamamatsu M64 \gls{MAPMT} is shown in \fig\ref{fig:Spectrum_EC107-PMTC-Pixel038}.
This spectrum was taken using the setup and procedure
described in the previous sections, and will be used here as a concrete example of the efficiency measurement technique.

The single photoelectron spectrum in \fig\ref{fig:Spectrum_EC107-PMTC-Pixel038} is a histogram of the charge received by the \gls{QDC} in each of one million gates, with the abscissa delimited in \gls{QDC} return codes (channels). 
The pedestal can be seen on the left side of the figure.
The threshold, shown in the figure by the horizontal line, is set in the valley between the single photoelectron peak and the pedestal at 2048 channels.
The peak to valley ratio of this pixel is 2.8, indicating a good photon counting resolution for this \gls{PMT}.
 
With the spectrum separated at the valley, the mean of the pedestal is at 2024.7 channels, shown in the figure by the upwards-facing triangle. 
Likewise, the mean of the single photoelectron peak, located at 2074.6 channels, is denoted by the downwards-facing triangle.
This gives a single photoelectron gain of 49.9 channels. In order to convert this relative gain into an absolute result the response of the \gls{QDC} must be taken into account.

The \gls{QDC} has been calibrated to check its linearity and to measure the conversion between
input charge and the QDC channels. The resulting conversion for the channel of the QDC used in this measurement was $19.3~\pm~0.2~$fC/channel,
giving a single photoelectron gain of $6.02~\pm~0.06~10^{6}$ for this pixel. The 1\% uncertainty on the absolute gain is dominated by the uncertainty on the response of the \gls{QDC}.

Counting each event above the threshold gives 12666 single photoelectrons collected in this spectrum.
Using the Poisson statistics, \eq\ref{eq:SPEcondition}), the contamination of two photoelectrons is 0.6\%. This could be treated as a correction, but here
it is taken as a systematic under-counting and is included in the systematic error. 
This is appropriate for a single photoelectron counting experiment where the number of photoelectrons is counted using a single discriminator, in which case the read out is not sensitive to 
higher numbers of photoelectrons.

A (monochromatic) LED with a wavelength of 378 nm was used as the light source, and the spectrum was taken over a total time of 503.9 seconds, as measured by the clock of the \gls{DAQ}.
The average power measured by the NIST photodiode during this time was $813.7~\pm~0.1~$pW. 
The attenuation of the integrating sphere and collimator used here were measured to be $6.874~\pm~0.003~10^{-8}$.  
% P(t), the power on the photodiode at each sample was also plotted to check that there were no spikes and in the LED output, etc.

Using \eq\ref{eq:EfficiencyCalculation}), the efficiency of this pixel using the chosen threshold is then $0.236~\pm~0.04$, an uncertainty of 1.9\%.
The uncertainty was estimated as in sect.~\ref{subsec:Single photoelectron efficiency}, but in this measurement the statistical error on the number of single photoelectron counts was not negligible. 
It is extremely important to note that this result for the efficiency and its corresponding uncertainty are for one position on the photocathode, one voltage supplied to the \gls{MAPMT}, and one counting threshold.

The result given is the absolute efficiency of this pixel averaged over the area illuminated by a light spot of $\approx0.07$~mm$^{2}$ (based on the collimator used in this measurement), 
and so the uniformity of the efficiency over the entire surface of the pixel remains to be taken into account.
This can be accomplished by measuring the efficiency at multiple locations within the pixel, and for this the precision X-Y movement of this setup is indispensable.  
If appropriate, the incidence angle dependence of the efficiency can also be studied. 
For the M64 \gls{MAPMT} the change in the efficiency was found to be on the order of 1\% between $-35$ and $+35$ degrees. 

The voltage supplied to the \gls{MAPMT} (here using a custom voltage division and 1100 V on the photocathode) determines the single photoelectron gain and the collection efficiency, and so any change in voltage will result in a variation of these properties. 
The efficiency is also strongly dependent on the chosen charge threshold, and so the translation of this result to another read-out system could be a source of significant 
systematic uncertainties (if the reference threshold is set by a DAC, for example).  The efficiency should thus be measured using the
power supply and read-out system of the experiment in question in order to benefit from the full precision of this method.

The relative efficiency within a single \gls{MAPMT} (or across several \glspl{PMT}) can be measured by illuminating the entire photocathode uniformly using the $\cos^{4}\theta$ dependence of the illumination from a Lambertian source.
This is achieved by placing the PMT(s) at a relatively large distance from the open port of the integrating sphere. At 30 cm, the non-uniformity with the setup used here is less than 1\% across a full M64 MAPMT.

The advantage of this type of relative measurement is that it
accounts for the variation in sensitive area of each pixel within the \gls{MAPMT} and gives the efficiency averaged over the surface of each. 
However, uniform illumination is not well-suited to an absolute measurement of the efficiency, due to the larger uncertainty on the number of photons incident within the pixel being characterized compared to illumination with a light spot
contained entirely inside the pixel.  

% The sixty-four spectra of the same PMT used for the absolute measurement, taken under uniform illumination,
% are shown in \fig\ref{fig:64SpectraAnalyzed}. 
% This type of measurement is important as  each anode of a \gls{MAPMT} displays its own gain and efficiency, typically varying by about 10\% within one PMT.
% Appropriately combined, the two measurements give the absolute single photoelectron gain and efficiency of each pixel in the \gls{MAPMT}. 

% \begin{figure*}[p]
%   \centering
%   \includegraphics[width=1.0\textwidth]{EC107-PMTC-Allpixels_v2}
%   \caption[64 Spectra Analyzed]{The 64 spectra of same PMT as in \fig\ref{fig:Spectrum_EC107-PMTC-Pixel038}.
% Here the PMT has been illuminated uniformly in order to measure the relative efficiency of each pixel.
% Each of these spectra have been analyzed in the same manner as in Figs.~\ref{fig:ExampleSPEspectra} and \ref{fig:Spectrum_EC107-PMTC-Pixel038}.
% For each pixel, the location of the one photoelectron peak mean and the valley have been found and marked. The red line 
% shows an extrapolation of the single photoelectron peak below the valley. In several spectra the analysis results are not reliable due to bad connections, noise, or pixels with a low gain. 
% }
%   \label{fig:64SpectraAnalyzed}
%  \end{figure*}

This calibration technique can also be applied to \glspl{SiPM} with adaptions to account for their properties.
In particular, single photoelectron counting using the threshold method, i.e.\ counting with a 
discriminator, is less advantageous for \glspl{SiPM} than for \glspl{PMT}, as the high resolution of \glspl{SiPM} allows each photoelectron peak to be resolved. 
This has the implication that
\begin{inparaenum}[i\upshape)]                                                                                                                                                                                                   
\item there is no need to work strictly in single photoelectron mode, and 
\item a \gls{QDC} read-out is more beneficial.    
\end{inparaenum}
Owing to the first point, the attenuation required to match the gain of a \gls{SiPM} to that of the NIST photodiode is much lower. This allows the calibration method to be more easily applied.
In addition, the statistical uncertainty on the \gls{SiPM} efficiency can easily be made negligible as there is no need for only 1\% of all events to be signal as when taking a single photoelectron spectrum.
On the other hand, the front-end electronics of an experiment using \glspl{SiPM} should be relatively more complex, including either a \gls{QDC} or an integrating preamplifier combined with an analog-to-digital convertor
in order to take advantage of the good photoelectron resolution of \gls{SiPM}.

\section{Conclusion}
In any application which requires that the absolute number of photons detected be known a precise knowledge of the efficiency of the photodetector in use is essential. 
% Examples in high-energy physics include the detection of ultra-high energy cosmic rays through the air fluorescence technique.
% The work presented is motivated by the need to calibrate the focal surface of the EUSO-Balloon instrument, which is a technical pathfinder for the future JEM-EUSO mission.
Here a complete discussion of photomultiplier calibration was presented with an emphasis on the difficulties and subtleties involved in a precision measurement of this type.
A technique for measuring the single photoelectron gain and efficiency of multi-anode photomultipliers with a total uncertainty of 1.7\% (1.6\% systematic) was presented. 
This precision is obtained by using a comparison to a NIST calibrated photodiode, and is an improvement over past methods, which typically give uncertainties on the order of 5 to 10\%.
This method can be applied to both vacuum photomultiplier tubes and other photodetectors, including measurements of the photo-detection efficiency (PDE) of \glspl{SiPM}.

The DAQ described in the sect.~\ref{sec:Measurement Setup} is a complete implementation of the absolute calibration technique for 64 pixel MAPMTs. 
This implementation includes not only CAMAC QDCs to read the anode signals of the 
MAPMT, but also a read-out of one or more photodiodes, and the control of a precision X-Y movement. This allows complete flexibility to perform complex measurements, such as scanning a \gls{MAPMT} photocathode pixel by pixel.
In addition, a complete analysis of all 64 channels is performed at the end of each run to reliably extract results from the measured single photoelectron spectra. All data are saved event by event, which makes complex analyses possible,
such as searching for coincidences between events in different pixels.

A complete measurement of the absolute efficiency of one Hamamatsu R11265-M64 \gls{MAPMT} was presented as an example of the application of the described characterization technique. 
The absolute efficiency in this example was measured with an uncertainty of 1.9\%, and this single result can be extended from the measured pixel to the entire \gls{MAPMT} 
with a final uncertainty of between 2 to 3\% through a relative measurement using uniform illumination. 
This setup has been used to perform a preliminary characterization of each \gls{MAPMT} in the EUSO-Balloon instrument, which will be reported in a future paper.

\section{Acknowledgments}
We wish to thank the CNES and IN2P3 for their support of the EUSO Balloon project, within which this work was undertaken.

\section{References}
%---bibliography

\end{document}